\documentclass[dvips,aps,rmp,twocolumn,groupedaddress]{revtex4}
\usepackage[sort&compress]{natbib}
\bibpunct{[}{]}{,}{a}{,}{,}
\usepackage{graphicx}
\usepackage{amsmath}
\usepackage{txfonts}
\usepackage{pxfonts}
\usepackage{pslatex}
\usepackage{textcomp}
\usepackage[T1]{fontenc}

\begin{document}
\title{Networks and Our Limited Information Horizon}
\author{M. Rosvall}\email{rosvall@tp.umu.se}
\affiliation{Department of Theoretical Physics, Ume{\aa} University,
901 87 Ume{\aa}, Sweden}
\author{K. Sneppen}
\affiliation{Niels Bohr Institute, Blegdamsvej 17, Dk 2100, Copenhagen, Denmark}
\homepage{http://cmol.nbi.dk}

\date{\today}

\begin{abstract}
In this paper we quantify our limited information horizon,
by measuring the information necessary to locate specific nodes in a network.
To investigate different ways to overcome this horizon,
and the interplay between communication and topology in social networks,
we let agents communicate in a model society.
Thereby they build a perception of the network
that they can use to create strategic links to improve their standing 
in the network.
We observe a narrow distribution of links when the
communication is low and a network with a broad distribution
of links when the communication is high.
\end{abstract}

\maketitle

\section{Introduction}
Communication is a fundamental element in maintaining the overall
cooperation between different parts of a complex system.
Because a complex system consists of many different parts,
it matters where signals are transmitted.
Thus signaling and traffic is in principle specific, 
with each message going from an unique sender to a specific target.
Networks are therefore a powerful way to represent this constrained 
communication of the real world \citep{rosvall}.

We start by using walks in networks with specific targets to quantify
the information necessary to locate specific nodes in the network \citep{hide-seek},
and also to investigate the constraints limited information sets on the navigability.
The process consists of extracting information at the nodes on the
walk between a source and a target. The subsequent question is therefore
the availability of this information.
We therefore let agents in a model society use local
communication to self-organize distant communication-pathways.
In this way we demonstrate that simple local rules allow agents
to build a perception of a dynamic system.
This perception guides a targeted signal across the network
beyond the information horizon \citep{friedkin,valverde1,horizon}.
Further we in this minimalistic model find that messages are most effectively
forwarded in the presence of hubs with funneling \citep{milgram1969},
like in scale-free networks, while transmission in hub-free networks
is more robust against misinformation and failures.

With the locally generated global information of the network, we
can take the model society one step further and let the agents
use the information to create strategic links.
In this way we are able to model the self-organization between communication
and topology in social networks, with a feedback between different
communication habits and the topology.
We observe a narrow distribution of links when the
communication is low and a system with a broad distribution
of links when the communication is high.

\section{Navigation in networks}
\begin{figure}[htp]
\centering
\includegraphics[width=0.95\columnwidth]{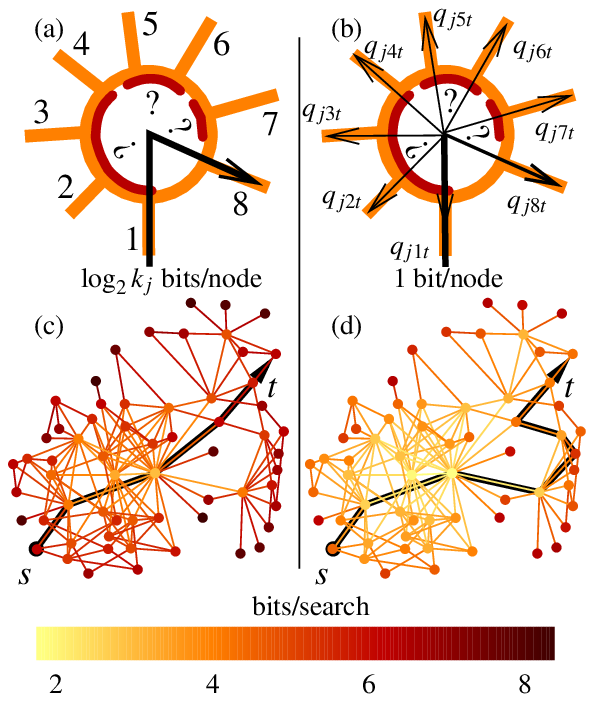}
\caption{\label{fig1}Navigation in networks. (a) It is possible to ask yes-no questions and
successively eliminate groups of wrong exits if the links are ordered.
Every yes-no question optimally reduces the number of possible links with $1/2$ and
the cost is $\log_2k$ to find the correct exit link. (b) The outcome of the elimination process is uncertain if the information that can be received is limited, and there is a finite probability to make mistakes ($q_{jit}$ in the direction $i$ at this node $j$ on the way to target node $t$). The color of the nodes and links in the lower panel represent the average number of questions necessary to navigate to the node. (c) With in total 13 bits it is possible to find the shortest path between the marked source and target. (d) With the information limited to 1 bit/node the navigation becomes here 3 steps longer (total cost is 8 bits).}
\end{figure} 

The simplest walker is the random walker,
which has been used to characterize topological features of
networks \citep{bilke,monasson}, including first passage times \citep{noh},
large scale modular features \citep{eriksen}, and
search using topological features \citep{adamic}.
Here we instead take the opposite approach and consider a direct walker.
In particular we quantify the information necessary
to locate specific nodes in the network \citep{hide-seek},
and investigate the constraints limited information sets on the navigability \citep{bit}.

A walk consists of stepping from node to node via
the links between them. The walk from a source node $s$ to a target node $t$ may be more
or less directed depending on the walkers ability to choose exit links that
lead it closer to the target (see Fig.\ \ref{fig1}(c-d)).
We first quantify the information cost in number of bits 
$I(s \to t) = \sum_{j \in \mathrm{path}(st)} \imath_{jt}$
it takes to navigate the shortest path from node $s$ to node $t$,
as the sum of the local information $\imath_{jt}$ on every node $j$
on a walk ``$\mathrm{path}(st)$'' leading to target $t$. That is, $\imath_{jt}$ is the number of
bits one needs on node $j$ to select one exit that leads to $t$.
If no degenerate paths exist, as in Fig.\ \ref{fig1}(a), then
$\imath_{jt} = \log_2 k_j$, where $k_j$ is the degree (number of links)
of node $j$, since the task is to select one link among $k_j$.

When there are two or more degenerate paths from $j$ to $t$,
the required information depends on the relative probabilities
that one wants to choose each shortest path with, and
$\imath_{jt}$ above generalizes to
\begin{equation}
\imath_{jt}=\log_2(k_j)+\sum_{i}q_{jit}\log_2 q_{jit},
\label{degen}
\end{equation}
where $q_{jit}$ is the
probability to choose a link to node $i$ from node
$j$ on a walk to node $t$ ($\sum_{i}q_{jit}=1$).
$q_{jit}=0$ if the link is not on the shortest path between $j$ and $t$.
We will choose the probability to leave a node along a link on a shortest path
between $s$ and $t$ to minimize the total information
cost $I(s \to t)$. Thus, if there are many degenerate paths,
the probability to exit to node $e$ from node $j$ on the shortest path to $t$ is
\begin{equation}
q_{jet}=\frac{p_{jet}}{\sum_{i} p_{jit}},\;\;
\mathrm{where}\;\; p_{jit}= \sum_{\mathrm{path}(it)} \;\;
\prod_{l \in \mathrm{path}(jit)} \frac{1}{k_l}
\label{qvalues}
\end{equation}
is the probability to walk the shortest path to $t$
from node $j$ via the link to node $i$ in an unbiased walk.

We now turn to the limited information perspective, 
and assume that the amount of information at a node is limited to $\imath$ bits
(see Fig.\ \ref{fig1}(b) and Java applet \citep{URLjavabit}).
The walk can now be substantially longer than
the actual shortest path. In Fig.\ \ref{fig1}(d) $\imath = 1$ and the 
walk is about 3 steps longer than in Fig.\ \ref{fig1}(c) with unlimited information.
To limit $\imath_{jt}$ to $\imath$
we blur the $q$-values of node $j$ in Eq.\ (\ref{qvalues}) 
by a $\epsilon_{jt} \in [0,\infty]$,
through $q_{jit} \rightarrow q_{jit}(\epsilon_{jt})=
\frac{q_{jit}+\epsilon_{jt}}{1+k_j \epsilon_{jt}}$
with $\epsilon_{jt}$ determined by
\begin{equation}
\imath_{jt}=\log_2(k_j)+\sum_{i}q_{jit}\log_2 q_{jit} \le \imath.
\end{equation}

\begin{figure}[htp]
\centering
\includegraphics[width=0.8\columnwidth]{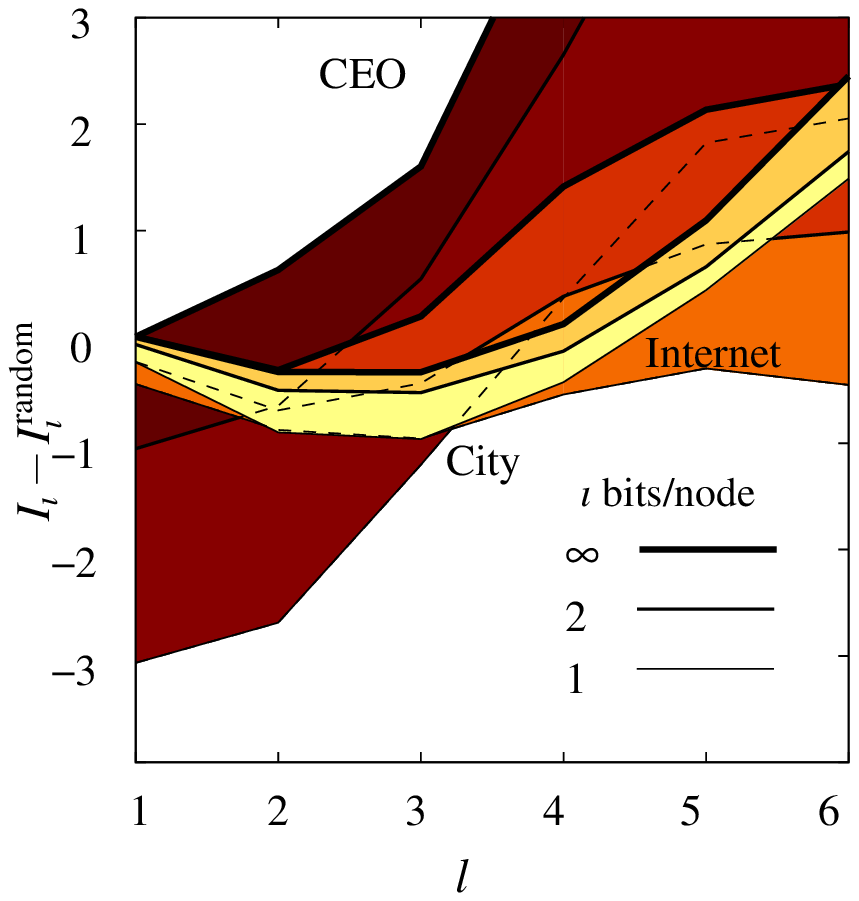}
\caption{\label{fig2}Information horizon in three real-world
networks. We compare information associated to navigation between nodes at distance $l$, with
the navigation in randomized counterparts (keeping the degree sequence) for
$\imath=\infty, \; 2$ and $1$. The Internet (hardwired Internet
of autonomous systems \citep{URLinternet}) is more sensitive to limited information
than the similarly sized CEO (Chief Executive Officers connected by links if
they sit at the same board \citep{ceo}).
The city network is the Swedish city Malm{\"o} with streets mapped to nodes and
intersections mapped to links \citep{city}. }
\end{figure} 

The navigability of a network is determined by its topology,
hence it depends on both the degree distribution and how
nodes of various degrees are connected to each other.
Here we focus on comparing a given real-world network with
its randomized counterparts, created
by rewiring links such that all nodes conserve their degree, and such that the
network remains globally connected \citep{maslov2002}.
In Fig.\ \ref{fig2} we resolve $I_{\imath}$
into $I_{\imath}(l)$ and examine the average information associated to 
walking to a specific node a distance $l$ away in the real and in the randomized ($I_{\imath}^{\mathrm{random}}(l)$) network \citep{horizon}.

The pattern that the real networks demand less information than their
randomized counterparts on short distance, $l<3$, suggests that many real-world networks favor
communication on short distance at the cost of constraining communication beyond this horizon.
Furthermore, this feature is more evident with limited than with complete information.

All results until now are based on the assumption that information is available
at every node, although the information can be limited.
To address the question of how this information can be assembled, we in the next section
turn to a social game and let agents in a model society use local
communication to build a global perception of the network.

\section{Self-assembly of information in networks}

\begin{figure}[htp]
\centering
\includegraphics[width=0.8\columnwidth]{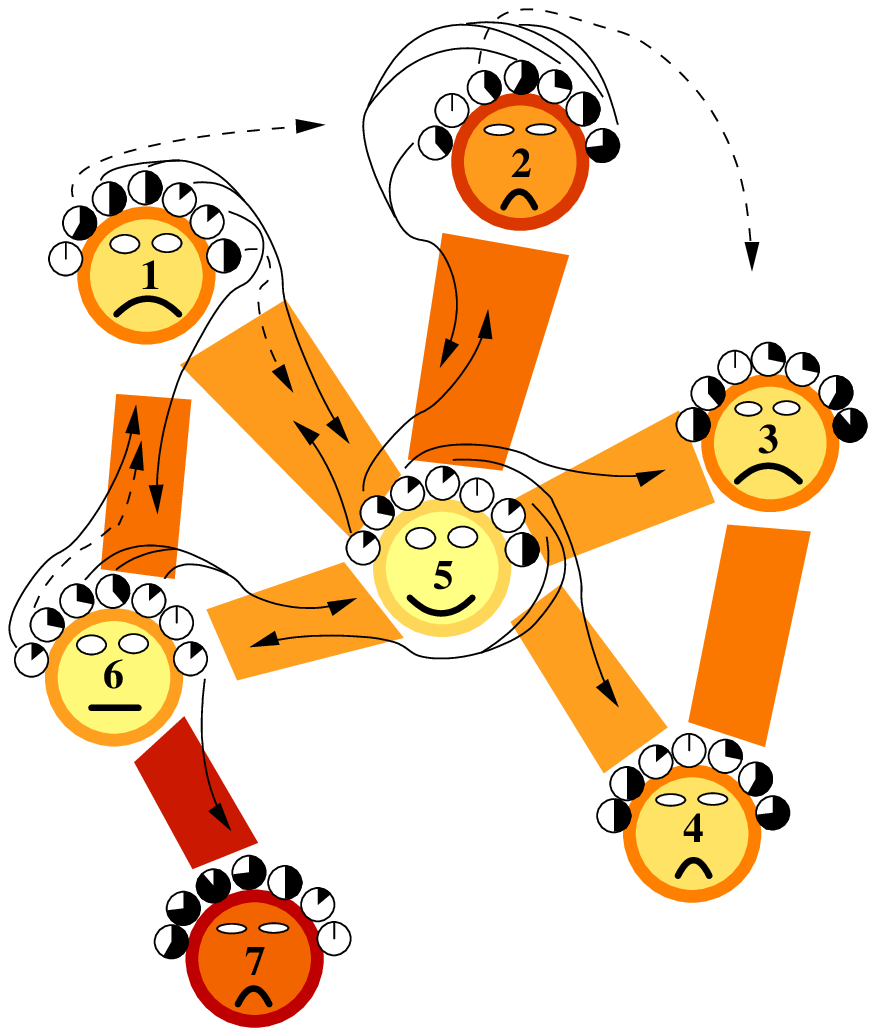}
\caption{\label{fig3}Modeling self-assembly of information in networks. Agents at nodes, connected by links, communicate with their connected acquaintances about any third target agent
in the network, and estimate the quality of the information by its age (clocks over heads correspond to, from left to right, the age of the information about agent \textbf{1}, \textbf{2},\dots \textbf{7}). The pointers are, for every agent, the acquaintance that connects most
efficiently to each of the other agents in the system (here only indicated for agent \textbf{1}, \textbf{2}, \textbf{5}, and \textbf{6}, dashed pointers are outdated).}
\end{figure}

To visualize our basic approach we illustrate in Fig.\ \ref{fig3}
a network composed of individual agents, each
of them connected to one or more acquaintances. Each individual communicates 
with its immediate neighbors to exchange information about agents
in other parts of the system. In this way every individual gradually builds up a
global perception by knowing people through people \citep{friedkin-infoflow}.
In our minimalistic model,
we allow each agent to have the information 
about which of its neighbors that connects most
efficiently to each of the other agents in the system.
Thus, a perfectly informed agent knows in which direction to send a 
message to any other agent in the system. If all agents were perfectly informed,
any message would be reliably forwarded from sender to recipient, using the
information of the subsequent agents along its paths \citep{milgram1969}.

The key question is how different communication rules of the
agents influence their possibility to obtain a reliable perception
that is robust to dynamical changes of the network.
Obviously, the agents need some index of quality that let
them judge whom of their acquaintaces that has the best knowledge of a particular agent.
We have found that the age of the information about an agent gives a very good estimate of the quality \citep{infoflow}.
That is, the \emph{perception} consists of, for every agent about any other agent:
\begin{itemize}
 \item The age of the information about the other agent \\(clocks in Fig.\ \ref{fig3}).
 \item From whom the information came \\(pointers in Fig.\ \ref{fig3}).
\end{itemize}
This defines the model together with the \emph{communication} event:
\begin{itemize}
\item
Select a random link and let the two agents that it connects communicate about a 
random third agent. The two agents also update their information about each other.
\end{itemize}
For example, when an acquaintance of agent \textbf{5} in Fig.\ \ref{fig3} obtains information about \textbf{5}, it
sets its pointer to \textbf{5}, and the information starts aging.
With successive communication events,
the information spreads from agent to agent and gets older and older
(we increase the age of all information when all links on average have participated in one communication event).
When two agents compare the validity of their pointers to a target agent,
like \textbf{1} and \textbf{6} to \textbf{5} in Fig.\ \ref{fig3},
they validate the newest information as the most correct one.
The agent with the oldest information changes its associated pointer to
its acquaintance in the communication event, and updates the clock.
By letting the agents memorize the acquaintances that provided the newest information 
about other agents together with the age of this information,
they will point in the direction of the fastest communication path from a target.
Moreover, the fastest communication paths are typically close to the shortest paths \citep{infoflow}.

\begin{figure}[htp]
\centering
\includegraphics[width=0.8\columnwidth]{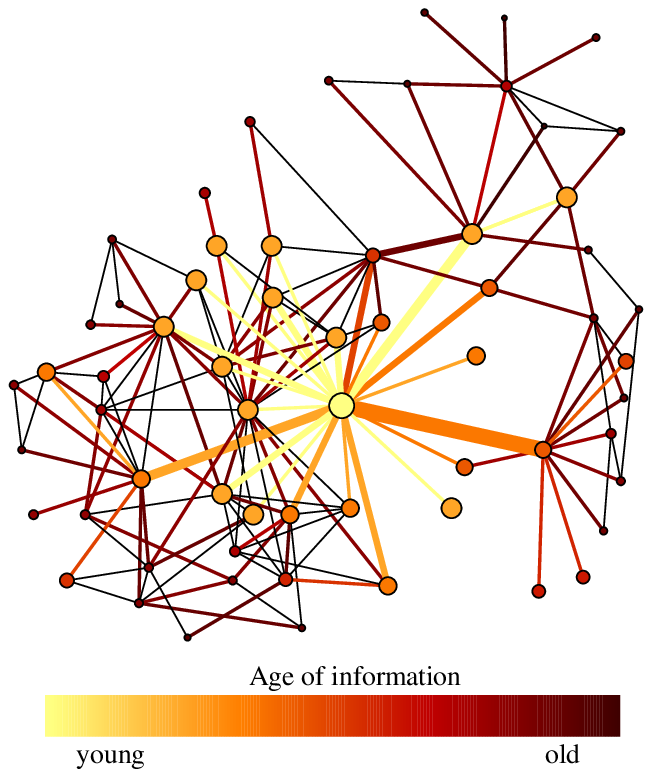}
\caption{\label{fig4}Self-assembly of information in networks. The size and the color of the nodes reflect the age of the information the well connected agent in the middle has of other agents. The width of the links reflects the relative amount of information they transfer to this agent, and the color the average quality (age) of this information. The agents make use of the hubs to create short communication-paths.}
\end{figure} 

Figure \ref{fig4} shows the perception around the central node in a model network (see also Java applet \citep{URLjavainfoflow}).
Clearly the information is most up to date in the immediate neighborhood 
of the agent, but that distant communication-pathways extend the whole network.
In a more detailed investigation, we found that messages are most effectively
forwarded in the presence of hubs with funneling \citep{milgram1969},
like in scale-free networks, while transmission in hub-free networks
is more robust against misinformation and failures.
This is in overall accordance with Stanley Milgram's famous experiment, where letters
were transmitted by sequences of acquaintance-acquaintance contacts across USA \citep{milgram,milgram1969}.
The choice of acquaintances was based on the participants' network perception, including also
geographic closeness of the acquaintance to the target (first steps)
and similarity of occupation (later steps) to forward messages \citep{killworth}.
Of course, these are two of many layers that could be added to the model.
However, the minimalistic model demonstrates that simple local rules allow agents
to build a perception of the system, which is enough to
overcome the information horizon \citep{friedkin} set by immediate acquaintances.
In this way the ``small world'' is really small \citep{milgram,kochen}, and it makes sense to talk
about navigation or search in networks
and to quantify the information associated to this process.

Given the agents' perception of the network it is tempting to take this social
game one step further, and in the next section we give the agents a social mobility.
The agents can thereby get new acquaintances to meet different interests.

\section{Self-organization of networks}

Social mobility may be seen as the response to the quest for better information access in a social system.
We let agents communicate to build a perception of a network as in the previous section,
and further allow the agents to use this information to create strategic links.
In this way we are able to investigate the feedback between different communication habits and the topology,
while the agents self-organize the social network.

The core of the model is the same as in the previous section.
To this we add the possibility to rewire the network, and the model can be formulated
in the two independent events \citep{inforew}:

\begin{itemize}
\item
\emph{Communication:}
Select a random link and let the two agents that it connects communicate about a 
random third agent. The two agents also update their information about each other.

\item
\emph{Rewiring:}
Select a random agent and let it use the local information to
ask an acquaintance about whom to establish a link to,
to shorten its distance to a randomly chosen other agent (the answer is the agent that the acquaintance points to).
Subsequently a random agent loses one of its links.
\end{itemize}
The communication event is typically repeated of the order of
number of links in the system for each rewiring event.

\begin{figure}[htp]
\centering
\includegraphics[width=0.9\columnwidth]{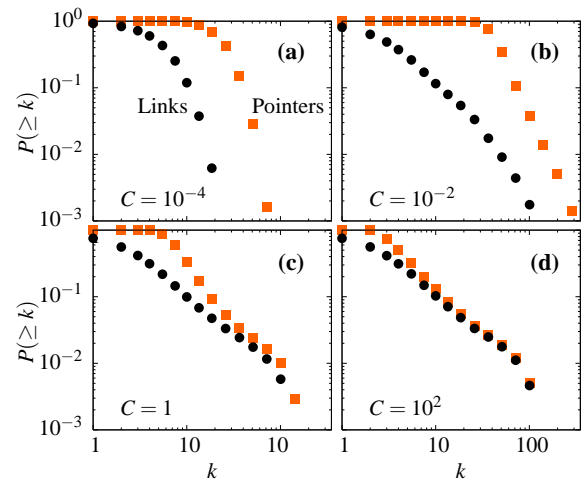}
\caption{\label{fig5}Illustration of the feedback from communication on the topology
of both the communication network and the perception network at four different levels of communication $C$.
$C=1$ corresponds to on average 1 communication event per link and rewiring event.
The system size is $N=1000$ agents connected by $L=2500$ links.}
\end{figure} 

We quantify the self-organization between communication,
the network, and the perception of this network, in Fig.\ \ref{fig5} (see also Java applet \citep{URLjavainforew}).
The perception network is defined by nodes as agents, and links between all pairs of nodes
where at least one of the corresponding agents has a pointer to the other agent.
Hence the perception network reflects the imperfect picture the agents have of their surrounding,
as outdated pointers cause diverging communication and perception networks.

We started with random Erd{\H o}s-Rényi networks \citep{erdos} (the results are independent of initial conditions)
and let the system evolve at different communication levels $C$.
The system size was $N=1000$ agents and $L=2500$ links.
$C \cdot L$ is the number of communication events per rewiring event in the network,
and the degree $k$ of a node is its number of links.
At low communication level, $C<1$,
the perception network has many more links than the communication
network, reflecting the failure of agents to perceive connections that are lost recently.
As $C$ approaches  $C\sim 1$
the perception network prunes its links whereas the communication network 
develops nodes with high degrees (the distribution can be approximated by a power law $P(k) \propto k^{-2.2}$).
At even higher values of $C$ the two networks converge toward the same broad degree-distribution---local communication gives rise to global organization.

The presented model describes a social game where the aim is to be central,
and a winner is an agent with many connections
that provide short and reliable communication to other agents.
The fact that we observe agents with a wide range of degrees
reflects the diversity of the possible outcomes of the game,
and raises the questions about whether there are some particular
strategies with which agents can improve their standing 
in the network? In a more detailed investigation, we found that individual increase of communication
gives both a local gain for the agents that adopt the communication strategy,
and a global gain for the whole system \citep{inforew}.

\section{Summary}

We started by investigating how the network topology affects the communication
ability in various networks,
and demonstrated that many real-world networks favors 
communication on short distance at the cost of constraining 
communication on long distance.
Thereafter we examined the ability to self-organize locally available information in a system
such that messages can be guided between distant parts of the network.
Our approach was to let agents chat in a model system
to self-organize distant communication-pathways.
We demonstrated that simple local rules allow agents
to build a perception of the system that is robust
to dynamical changes and mistakes.

Finally we explored the local dynamic origin of global network organization
by modeling the response to information transfer in a simplified social system.
We found that a low communication level results in chaotic or Erd{\H o}s-Rényi like networks,
whereas higher communication levels with more reliable communication-pathways
lead to structured network topologies with broad degree-distributions.

Overall we have used networks to quantify our limited information horizon.
In a society the information horizon is set by each individual's social contacts,
which in turn is a part of the global network of human communication.
By measuring the information necessary to locate specific nodes in a network
we were able to quantify this horizon, and by introducing a social game we
investigated different ways to overcome the horizon.
\begin{quotation}
\noindent
We acknowledge the support of The Danish National Research Foundation:
``Models of Life'' at NBI.
\end{quotation}

\end{document}